\documentclass[showpacs,aps,pra,twocolumn]{revtex4-1}
\usepackage{amsfonts,hyperref}
\usepackage{graphicx}
\usepackage{amssymb}
\usepackage{epsfig}
\usepackage{amsmath}

\begin{document}

\title{Quantum Transport of Bosonic Cold Atoms in Double Well Optical
Lattices}
\author{Yinyin Qian}
\author{Ming Gong}
\author{Chuanwei Zhang}
\thanks{Author to whom correspondence should be addressed; Email:
cwzhang@wsu.edu}

\begin{abstract}
We numerically investigate, using the time evolving block decimation
algorithm, the quantum transport of ultra-cold bosonic atoms in a double
well optical lattice through slow and periodic modulation of the lattice
parameters (intra- and inter-well tunneling, chemical potential, \textit{etc.%
}). The transport of atoms does not depend on the rate of change of the
parameters (as along as the change is slow) and can distribute atoms in
optical lattices at the quantized level without involving external forces.
The transport of atoms depends on the atom filling in each double well and
the interaction between atoms. In the strongly interacting region, the
bosonic atoms share the same transport properties as non-interacting
fermions with quantized transport at the half filling and no atom transport
at the integer filling. In the weakly interacting region, the number of the
transported atoms is proportional to the atom filling. We show the signature
of the quantum transport from the momentum distribution of atoms that can
measured in the time of flight image. A semiclassical transport model is
developed to explain the numerically observed transport of bosonic atoms in
the non-interacting and strongly interacting limits. The scheme may serve as
an quantized battery for atomtronics applications.

\end{abstract}
\affiliation{Department of Physics and Astronomy, Washington State
University, Pullman, WA, 99164 USA}
\pacs{03.75.Lm, 03.65.Vf}
\maketitle


\section{Introduction}

Quantum charge pumping, a coherent quantum transport process that generates
steady charge currents of electrons through adiabatically and periodically
time-varying potentials, is a standard method for the charge transport in
solid state circuits \cite{Thouless,Niu1,pump1,pump2,Xiao}. In the quantized
charge transport, the number of particles pumped out during each cycle of
the potential modulation is an integer and can be understood using a Berry
phase \cite{Berry}. Because of the precise control on the amount of pumped
charges at the single electron level, the quantum charge pumping has found
important applications in many electronic devices \cite{app1,app2}. It also
provides a solid foundation to the modern theory of electric polarization
\cite{polar1,polar2,polar3}. Similar ideas have also been extended to
another degree of freedom of electrons, the spin, leading to the quantum
spin pumping that plays a crucial role in spintronics \cite%
{spinpump1,spinpump2,spinpump3}. However, the exact quantized charge
transport is usually difficult to observe in experiments because of
unavoidable complexity due to impurities, disorder and interactions in the
solid.

In view of the significance of electronics and spintronics, there has been a
great deal of interest recently for developing a one-to-one analog of
complex and interesting electronic materials, circuits, and devices using
ultra-cold neutral atoms \cite{at1,at2,at3}. This field, known as
atomtronics, is a significant extension of the recent great efforts on
emulating condensed matter physics using ultra-cold atoms \cite%
{ole1,ole2,ole3}. Important concepts such as atomic batteries, diodes, and
transistors, have been proposed recently for atomtronics. It would be
natural and also important to investigate the coherent quantum transport
process of cold atoms in optical lattices, which is an important element of
atomtronic devices. A straightforward method for the transport, of course,
is by applying an external force. However, using external forces not only
involves the acceleration of atoms that may change the motional states of
atoms, but also may be limited by the Bloch oscillation of atoms in optical
lattices. Recently, a lot of attention has been focused on the quantum
transport of atoms in optical lattices without involving external forces
\cite{oriol07,creffield}. For instance, it has been proposed that quantum
transport of atoms may be achieved in optical double well lattices through
the fast periodic modulation (in a non-adiabatic manner) of the inter- and
intra-well tunnelings \cite{oriol07}\textbf{. }It was also proposed \cite%
{creffield} that a fast oscillating linear potential in optical lattices can
yield the quantum ratchet effect due to coherent destruction of tunneling
\cite{grossmann}, leading to quantum transport of atoms. However, all these
schemes require accurate control of the lattice parameters and their time
variation.

In this paper, we study the quantum transport of ultra-cold bosonic atoms in
optical lattices without involving the accurate control of the lattice
parameters and their time dependence. We consider atoms that are initially
prepared in certain region of a double well optical lattice \cite%
{doublewell1,doublewell2,doublewell3,doublewell4,doublewell5,doublewell6,
doublewell7,doublewell8,doublewell9,Rousseau,Schlagheck}. We show that
quantized number of atoms can be transferred to another region of the
lattice through periodic and slow modulation of the optical lattice
parameters (intra- and inter-well tunneling, chemical potential, \textit{etc.%
}). Such quantized atom pumping may serve as a quantum atom battery for
atomtronic devices. Note that there is an important difference between the
atom transport in optical lattices and the charge transport in the solid.
While the electrons in the solid are fermions, cold atoms in optical
lattices can be bosons, therefore the Bose-Einstein statistics, instead of
the Fermi-Dirac statistics, governs the transport dynamics. This important
difference will be illustrated in this paper\textbf{.} Another important
difference is that while electrons are periodically distributed in crystals,
the density distribution of atoms can be prepared locally in the optical
lattice and transferred to another region through the quantum transport
process with the periodic modulation of the lattice parameters. We emphasize
that although the periodic modulation of the lattice parameters in the
parameter space are needed in the transport process, the rate of change of
the parameters, the initial and final values of each parameter during its
variation do not need to be accurately controlled. This is because the
transport process depends on the topology of the loop in the parameter
space, instead of the exact parameter loop or the rate of change of the
parameters along the loop. Therefore the quantum transport process is very
robust against errors in the parameter modulation.

We find the transport of bosonic atoms depends strongly on the atom filling
per double well as well as the interaction. The strongly interacting bosonic
atoms behave similarly as the non-interacting fermions with quantized
transport at the half filling and no mass transport at the integer filling.
In the weakly interacting region, the number of transported atoms is
proportional to the atom filling. The investigation is based on the
numerical simulation of the exact quantum dynamics of cold atoms in double
well optical lattices using the time evolving block decimation (TEBD)
algorithm \cite{Vidal,TEBD}. The transport properties observed in the
numerical simulation are also understood by developing an analytical theory
using a semiclassical transport model in the non-interacting and strongly
interacting limits.

\begin{figure}[t]
\includegraphics[width=1.0\linewidth]{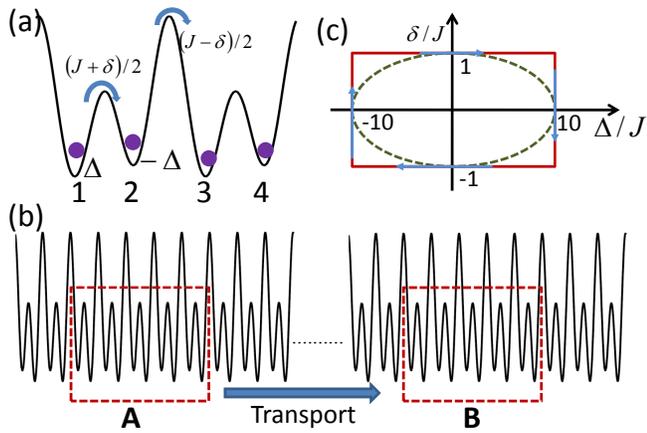} \vspace{-0pt}
\caption{(Color online) Schematic plot of the quantum transport of
ultra-cold bosonic atoms in a double well optical lattice. (a) Illustration
of the double well optical lattice parameters. (b) Illustration of \ the
quantum transport process of bosonic atoms from region A to region B in the
lattice. (c) The periodic modulation of the lattice parameters for the
quantum transport process. Two different types of loops in the parameter
space are considered: square (solid line) and elliptical (dashed line).}
\end{figure}

The paper is organized as follows: Section II describes the physical system:
the cold bosonic atoms in a double well optical lattice. Section III
presents the numerical results on the transport of bosonic atoms in the
optical lattices through the periodic modulation the lattice parameters. In
Section IV, we provide a semiclassical transport model to explain the
numerically observed transport of bosonic cold atoms in the non-interacting
and strongly interacting limits. Section V consists of conclusions.

\section{Cold Bosons in double well optical lattices}

Consider ultra-cold bosonic atoms confined in a one dimensional double well
optical lattice (Fig. 1a). An optical lattice is a standing wave of coherent
off-resonance light created by the interference of two or more laser beams.
In experiments, the 1D double well optical lattice has been realized by
superimposing two laser beams with two different wavelengths $\lambda
_{2}=\lambda _{1}/2$ \cite{doublewell1,doublewell2,doublewell3,doublewell4},
or the Fourier synthesis of asymmetric optical potentials with spatial
periodicity $\lambda /2n$ ($n$ is integer) \cite{ritt}. In these
experiments, the dynamics along the other two dimensions are frozen to the
ground states by optical lattices with high potential depths ($\gg $ recoil
energy). With the standard tight-binding approximation, the dynamics of
atoms in the double well optical lattices can be described by a Bose-Hubbard
Hamiltonian \cite{BH}
\begin{eqnarray}
H &=&-\sum\limits_{j=1}^{N-1}\left[ \frac{J}{2}-\left( -1\right) ^{j}\frac{%
\delta }{2}\right] \left( a_{j}^{\dag }a_{j+1}+c.c.\right)  \notag \\
&&+\sum\limits_{j=1}^{N-1}\left[ V_{j}+\left( -1\right) ^{j+1}\Delta \right]
n_{j}+Un_{j}\left( n_{j}-1\right) ,  \label{Ham1}
\end{eqnarray}%
where $\left[ J-\left( -1\right) ^{j}\delta \right] /2$ describes the inter-
or intra-well tunneling, $V_{j}$ is the external potential (e.g., the
harmonic trap, disorder, \textit{etc}.), $2\Delta $ is the chemical
potential difference between two neighboring lattice sites in a double well,
and $U$ is the on-site interaction strength between atoms. In experiments,
the tunneling $\left[ J-\left( -1\right) ^{j}\delta \right] /2$ can be
adjusted by varying the intensities of the two laser beams, the chemical
potential $\Delta $ can be tuned by shifting one laser beam with respect to
the other, the on-site interaction $U$ between atoms can be changed using
the Feshbach resonance \cite{FR}, the disorder potential can be created
using optical speckle potentials \cite{disorder}.

\begin{figure}[t]
\includegraphics[width=0.8\linewidth]{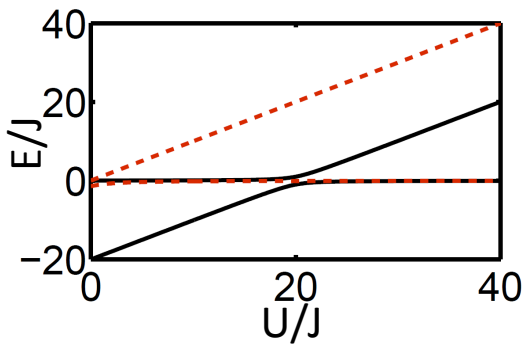} \vspace{-0pt}
\caption{(Color online) Plot of the lowest two eigenenergies with respect to
$U$ for two atoms in a double well. Solid line: $\Delta =0$. Dashed line: $%
\Delta =10J$. }
\end{figure}

The transport process of cold atoms is illustrated in Fig. 1b. Initially,
the bosonic atoms are confined within a group of lattice sites (denoted as
region A) using a harmonic or box trapping potential \cite{Bloch}, which
will be removed after the transport starts. The transport is realized by
moving the atoms in the region A to another group of lattice sites (denoted
as region B) by modulating the lattice parameters periodically, but without
the actual movement of the lattice. Therefore it is a quantum transport
process through the tunneling dynamics. To simplify the study, we assume a
uniform distribution of the atoms in the region A at the time $t=0$. During
the transport process, the lattice parameters $\delta $ and $\Delta $ are
tuned along a periodic close loop in the parameter space, as illustrated in
Fig. 1c.

The time-dependent dynamics of atoms in the optical lattice are investigated
numerically using the TEBD algorithm \cite{Vidal,TEBD}. The TEBD algorithm
is an effective method for simulating the exact one-dimensional quantum
dynamics when the quantum entanglement of the system is low. It works well
for the Hamiltonian (\ref{Ham1}) where the only non-local terms are the
nearest neighboring tunneling. To check the validity of our numerical
program, we have compared the results from the TEBD and that from the exact
diagonalization for a small lattice system and they agree with a high
accuracy. Henceforth, we set the energy unit of the system as $J$, and the
time unit as $J^{-1}$.

We assume the initial parameter $\delta /J=1$ to obtain a well-defined
localized initial state. With this parameter, the inter-well tunneling
vanishes and the initial wavefunction of the system is a product of the
wavefunction in each double well. We use the number states of atoms at each
lattice site as the basis states to represent the wavefunction. For
instance, consider a double well lattice with a lattice length $N=20$ and $%
\Delta /J=0$ initially. Ten atoms are uniformly distributed in the region A
between sites 5 and 14. In the region A, the wavefunction in the $l$-th
double well can be written as
\begin{equation}
\psi _{l}=c_{20}\left\vert 20\right\rangle _{l}+c_{02}\left\vert
02\right\rangle _{l}+c_{11}\left\vert 11\right\rangle _{l},  \label{wave1}
\end{equation}%
where $c_{20}=c_{02}=\sqrt{\frac{J^{2}}{\left( \sqrt{U^{2}+4J^{2}}+U\right)
\sqrt{U^{2}+4J^{2}}}}$, $c_{11}=\sqrt{\frac{\sqrt{U^{2}+4J^{2}}+U}{2\sqrt{%
U^{2}+4J^{2}}}}$. $\left\vert nm\right\rangle $ is the state with $n$ ($m$)
atoms in the left (right) well. Outside the region A, the wavefunction in
each double well is just $\phi _{l}=\left\vert 00\right\rangle _{l}$.

\begin{figure}[t]
\includegraphics[width=1.0\linewidth]{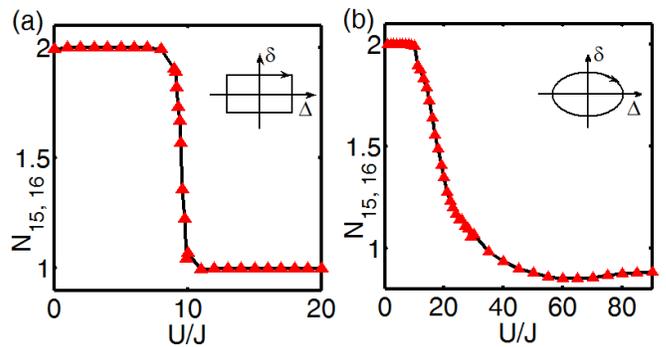} \vspace{-20pt}
\caption{(Color online) Plot of the number of atoms in the lattice sites 15
and 16 with respect to the atom interaction strength $U$ after one cycle of
the parameter modulation illustrusted in the insets. Initially ten atoms
uniformly occupy the lattice sites 5 to 14. (a) A square loop. (b) An
elliptical loop.}
\end{figure}

The quantum transport process are accomplished by modulating the lattice
parameters $\left( \delta ,\Delta \right) $ periodically in the parameter
space, as illustrated in Fig. 1c. Two different loops (square and
elliptical) are considered and their effects will be discussed later in the
paper. We choose the rate of the change of the parameters $\left( \delta
,\Delta \right) $ such that the tunneling process is adiabatic in each
double well and the atoms always stay on the ground state of the double
well. During the whole transport process, the atoms are assumed to be on the
motional ground state at each lattice site. This assumption can be realized
by using a double well lattice with an intra-well potential depth $V_{L}\sim
6E_{R}$, where $E_{R}=\frac{h^{2}}{2m\lambda ^{2}}$ is the atom recoil
energy, $h$ is the Planck constant, $\lambda $ is the wavelength of the
short wavelength laser, and $m$ is the atom mass. This potential depth
yields a tunneling rate $J\sim 0.05E_{R}$ \cite{Greiner}. Therefore the
required chemical potential shift $\Delta \sim 0.5E_{R}$ for the transport
process are much smaller than the energy gap between the ground and first
excited motional bands (typically several $E_{R}$).

\begin{figure}[b]
\includegraphics[width=1.0\linewidth]{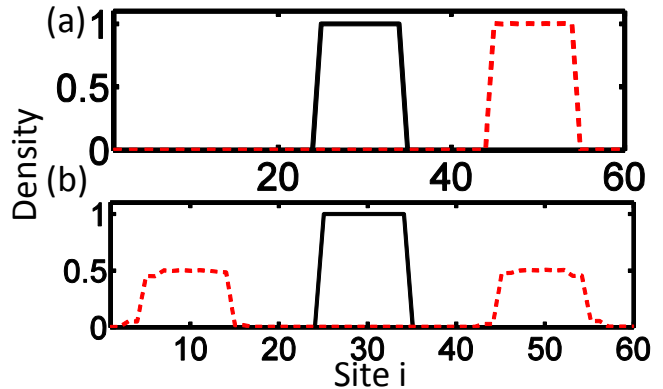} \vspace{0pt}
\caption{(Color online) Plot of the density distribution of atoms after 10
cycles of the parameter modulation with the square loop. Solid line: initial
density distribution. Dashed line: final density distribution. (a) $U=5J$.
(b) $U=20J$.}
\label{ten}
\end{figure}

In each isolated double well with two atoms, the Hamiltonian can be written
as%
\begin{equation}
H_{d}=\left(
\begin{array}{ccc}
2\Delta +U & 0 & -J \\
0 & -2\Delta +U & -J \\
-J & -J & 0%
\end{array}%
\right)  \label{Ham2}
\end{equation}%
on the basis $\left\{ \left\vert 20\right\rangle ,\left\vert 02\right\rangle
,\left\vert 11\right\rangle \right\} $. Without interaction between atoms ($%
U=0$), the eigenenergies are $E_{\pm }=\pm \sqrt{4\Delta ^{2}+2J^{2}}$, $%
E_{3}=0$. Therefore the energy gap between the ground state $E_{-}$ and the
first excited state $E_{3}$ is larger than $\sqrt{2}J$. In the presence of
strong interaction $U$ (neglect $\Delta $), the three eigenenergies are $%
E_{\pm }=\frac{1}{2}U\left( 1\pm \sqrt{1+8J^{2}/U^{2}}\right) $, $E_{3}=U$.
The energy gap between the ground state $E_{-}$ and the first excited state $%
E_{3}$ is $\sim U$. The crossover between different energy bands in Fig. 2
is essentially a Landau-Zener tunneling process. In both parameter regions,
the Landau-Zener adiabatic criteria in the double well can be easily
satisfied by requiring $\frac{d\Delta }{dt}=2\epsilon _{1}J^{2}$ along the
horizontal loop in Fig. 1c. Along the vertical loop, we choose $\frac{%
d\delta }{dt}=\epsilon _{2}\Delta ^{2}$ to avoid the tunneling between two
wells with different chemical potentials. Here $\epsilon _{1}=0.1\ll 1$, $%
\epsilon _{2}=0.01\ll 1$ are the adiabatic parameters. The energy gap is
also numerically calculated and plotted in Fig. 2 for general parameters. We
see the gap is always larger than $\sqrt{2}J$ for different $\Delta $ and $U$%
. When the lattice parameters are modulated along the elliptical loop in
Fig. 1c, there is no overall adiabaticity among double wells because atoms
can now diffuse to a long distance if a long time period is used for the
parameter modulation. Since the change of the parameters is slow, the cycle
frequency of the periodic lattice modulation is much smaller than the energy
gap.
\begin{figure}[b]
\includegraphics[width=1.0\linewidth]{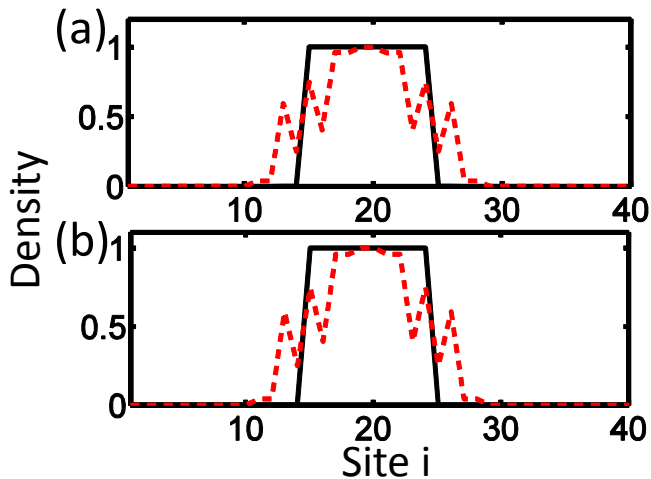} \vspace{-10pt}
\caption{(Color online) Plot of the diffusion of atoms after one cycle of
the parameter modulation. $\Delta $ is fixed at $0$. $\protect\delta $
varies from $J$ to $-J$ and then back to $J$. Solid line: initial density
distribution. Dashed line: final density distribution after one cycle. (a) $%
U=5J$. (b) $U=20J$. }
\label{diffusion}
\end{figure}

In the case of one atom per double well, the Hamiltonian in each double well
can be written as
\begin{equation}
H_{s}=\left(
\begin{array}{cc}
\Delta & -J \\
-J & -\Delta%
\end{array}%
\right)  \label{Ham3}
\end{equation}%
on the basis $\left\{ \left\vert 10\right\rangle ,\left\vert 01\right\rangle
\right\} $. The eigenenergies are $E_{\pm }=\pm \sqrt{\Delta ^{2}+J^{2}}$
and the minimum energy gap is $2J$, therefore the above rate of the change
of the lattice parameters still keeps the adiabaticity in each double well.

\begin{figure}[b]
\includegraphics[width=0.8\linewidth]{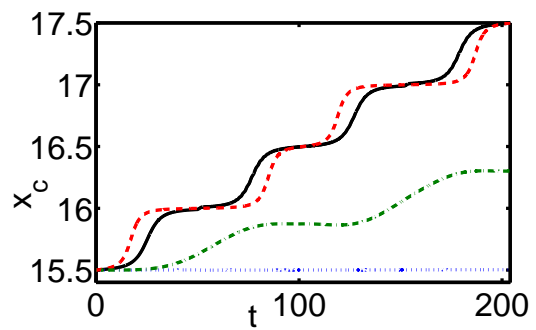} \vspace{-0pt}
\caption{(Color online) Plot of the center of mass motion of the atoms in
one cycle of the potential modulation. Initially six atoms are uniformly
distributed in three double wells. Solid line: $U=5J$, square loop. Dashed
line: $U=5J$, elliptical loop. Dotted line: $U=20J$, square loop. Dashed
dotted line: $U=20J$, elliptical loop.}
\label{COMsingle}
\end{figure}

\section{Quantum transport in double well optical lattices}

We numerically integrate the time-dependent Schr\"{o}dinger equation with
the Hamiltonian (\ref{Ham1}) using the TEBD algorithm and calculate the
density distribution of atoms at different lattice sites. In Fig. 3, we plot
the number of atoms at lattice sites 15 and 16 after one cycle of the
parameter modulation described in Fig. 1c. The lattice system contains total
$N=20$ sites with 10 atoms uniformly distributed between sites 5 and 14.
Clearly the number of atoms transported to the neighboring double well
(sites $15$ and $16$) depends on the interaction and the parameter
modulation loop. In the weak interaction region ($U<\Delta $), two atoms are
transferred to sites $\left( 15,16\right) $, which is the same as that for
non-interacting bosons. In the case of non-interacting bosons, atoms follow
the changes of the potential minima in the lattice system and it is easy to
see that atoms would like to move one double well distance (two lattice
sites) after one cycle of the parameter modulation. As the interaction
increases, there is a sharp transition for the square loop, where the number
of the transported atoms at sites 15 and 16 becomes one (Fig. 3a). In this
region, the atoms behave the same as non-interacting fermions, and two atoms
cannot occupy the same lattice site. The same physical picture applies to
more cycles of the parameter modulation. In Fig. 4, we plot the density
distribution of atoms after 10 cycles of the parameter modulation. For a
small interaction strength, the atoms are transported to the target region
that are 20 lattice sites away from the original region (Fig. 4a). For a
large interaction strength, atoms move to both left and right directions
(Fig. 4b), and the center of mass of the atoms does not change (\textit{i.e.}%
, no mass transport).

\begin{figure}[t]
\includegraphics[width=0.8\linewidth]{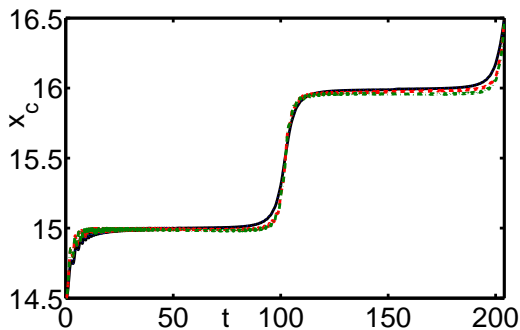} \vspace{-0pt}
\caption{(Color online) (a) Plot of the center of mass motion of the atoms
in one cycle of the potential modulation. Initially six atoms are uniformly
distributed in six double wells. Solid line: $U=5J$, square loop. Dashed
line: $U=5J$, elliptical loop. Dotted line: $U=20J$, square loop. Dashed
dotted line: $U=20J$, elliptical loop.}
\label{COMhalf}
\end{figure}

When the parameter modulation is along the elliptical loop, there is a
diffusion of the atoms along the lattices because the locally confined
initial state is not the ground state of the Hamiltonian. In Fig. 3b, we
plot of the number of atoms at lattice sites 15 and 16 after one cycle of
the parameter modulation along the elliptic loop. The main feature of Fig.
3a is still kept. In the weakly interacting region, the number of
transported atoms is two, but the sharp transition is smoothed out and the
number of atoms at the large $U$ is no longer one because of the diffusion
of the atoms through the lattice. In Fig. 5, we plot the density
distribution of atoms after one cycle of the parameter modulation with a
fixed $\Delta =0$ and varying $\delta $ between 1 and $-1$. The diffusion of
the atom density depends strongly on the rate of the change of the
parameters. To suppress such uncontrolled diffusion, henceforth we mainly
consider the square loop.

The atom transport in the optical lattice can also be described using the
center of the mass (COM) motion of the atoms. In Fig. 6, we plot the COM
motion of atoms in the double well lattice after one cycle of the parameter
modulation. We see in the weakly interacting region, the COM is shifted by
two lattice site for both elliptical and square loops. In the strongly
interacting region, the COM does not change for the square loop, indicating
no mass transport, but varies for the elliptical loop, showing the asymmetry
of the diffusion process in the parameter modulation process.

In the above discussion, we consider that there are two atoms per double
well (i.e., integer filling) in the initial state. Another interesting
region is the half filling, that is, one atom per double well. In Fig. 7, we
plot the COM motion of atoms after one cycle of the parameter modulation at
the half filling. In both weakly and strongly interacting regions, the shift
of the COM is two lattice sites, indicating that the non-interacting bosons
and fermions behaves similarly at the half filling although their transport
properties are completely different at the integer filling. The elliptical
and square loops also yield similar results.

\begin{figure}[t]
\includegraphics[width=0.8\linewidth]{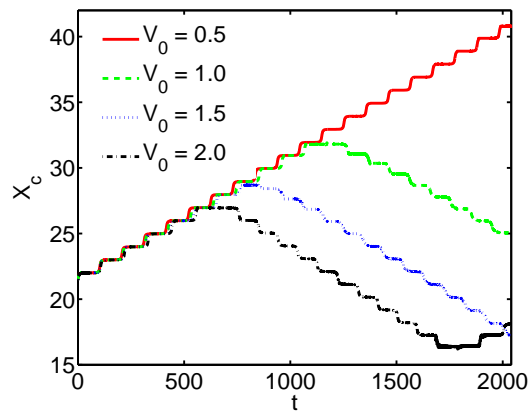} \vspace{-0pt}
\caption{(Color online) The transport of atoms in the presence of a harmonic
trapping potential $V_{i}=V_{0}\left( i-x_{o}\right) ^{2}$, where $x_{o}$
locates at the center of the lattice.}
\label{other}
\end{figure}

In addition to the double well optical lattice potential, the atoms may also
experience an overall harmonic trapping potential. In Fig. 8, we plot the
transport of atoms in the presence of an harmonic trap. Clearly the harmonic
trapping potential does not affect the transport even far away from the trap
center. More interestingly, the initial right-moving COM motion turns to the
left-moving motion after it reaches the maximum position that is determined
by the harmonic trapping frequency. In a realistic experiment, this
phenomenon corresponds to a dipole oscillation of atoms generated by the
periodic modulation of the lattice parameters. The turnaround can be
understood from the fact that when the COM reaches certain $x$, the chemical
potential difference $2\Delta $ cannot overcome the potential difference
between two wells induced by the harmonic trap. Therefore atoms do not move
at that cycle. Further modulation of the lattice parameters then provides a
driving of atoms to the left-moving direction.

\begin{figure}[b]
\includegraphics[width=1.0\linewidth]{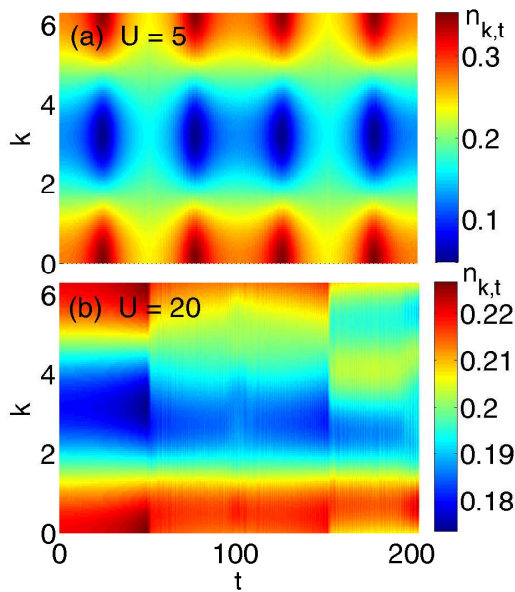} \vspace{-0pt}
\caption{Plot of the momentum distribution of atoms in the time of flight
image. (a) $U=5J$ and (b) $U=20J$.}
\label{fig-nk}
\end{figure}

Finally, we discuss how to observe the quantized atom transport in the
double well optical lattices. In principle, the transport dynamics can be
observed using the single atom detection technology demonstrated recently
\cite{singleatom1,singleatom2,singleatom3,zhang}. Here we consider the
signature of the quantum transport in more conventional experimental
techniques: the momentum distribution in the time-of-flight image. The
momentum distribution of the atoms in the optical lattices can be calculated
as \textbf{\ }%
\begin{equation}
n_{k,t}=\sum_{i,j}\langle \psi (t)|a_{i}^{\dagger }a_{j}|\psi (t)\rangle
e^{ik(i-j)}.
\end{equation}%
The numerical results are shown in Fig. \ref{fig-nk} for $U=5J$ (a) and $%
U=20J$ (b). In the weak-interacting case, the atoms in one double well are
perfectly transported to another double well and the final density
distribution just have a global shift of the position, there we expect $%
n_{k,t}=n_{k,t+T}$ after one period, as shown in Fig. \ref{fig-nk}a. In the
strong-interacting case, the cold atoms move along two opposite directions,
therefore $n_{k,t}\neq n_{k,t+T}$ and the coherent peaks in the momentum
distribution disappear. We find the momentum distribution is generally
independent of the loops used in TEBD calculations. In experiments, the
momentum distribution can be measured directly from the time-of-flight
image, thus provides a direct experimental signature for the quantum
transport.

\section{A semiclassical theory}

Physically, the numerical results obtained in the above section at the half
filling may be understood through the change of the Wannier function of
bosons in a periodic lattice due to the modulation of the lattice
parameters. The center of the Wannier function follows the potential minimum
of the double well, and can move from one double well to the neighboring one
in one cycle. In this section, we present a semiclassical theory for the
quantum transport of the non-interacting bosons in the double well lattice
to understand the numerical results observed in the above section. On the
other hand, the strongly interacting bosonic atoms are equivalent to
non-interacting fermionic atoms, therefore their transport properties are
the same as the non-interacting electrons. For simplicity, we consider a
periodic system to avoid the diffusion.

With a periodic boundary condition, we can transform the Hamiltonian (\ref%
{Ham1}) without the interaction ($U=0$) to the quasimomentum space
\begin{equation}
H=\Gamma \left( q\right) \cdot \sigma   \label{Ham4}
\end{equation}%
by using the Fourier transformation $a_{2j-1}=\sum_{q}e^{iq\cdot \left(
2j-1\right) a/2}a_{q\uparrow }$, $a_{2j}=\sum_{q}e^{iq\cdot
ja}a_{q\downarrow }$, where $\Gamma \left( q\right) =\left( -J\cos \frac{qa}{%
2},\delta \sin \frac{qa}{2},\Delta \right) $, and the spin up and down
correspond to the left and right sites of the double well. The Hamiltonian (%
\ref{Ham4}) has two energy bands $\alpha =\pm $ with the dispersion
\begin{equation}
\varepsilon _{\alpha }=\pm \sqrt{J^{2}\cos ^{2}\frac{qa}{2}+\delta ^{2}\sin
^{2}\frac{qa}{2}+\Delta ^{2}}  \label{energy}
\end{equation}

The velocity of an atom in the bands satisfies the semiclassical equation of
motion \cite{Niu3}%
\begin{equation}
\dot{x}_{\alpha }\left( q,t\right) =\frac{\partial \varepsilon _{\alpha }}{%
\partial q}-\Omega _{qt}^{\alpha },  \label{EOM}
\end{equation}%
where $\Omega _{qt}^{\alpha }=-2$Im$\left\langle \frac{\partial \Phi
_{\alpha }}{\partial q}|\frac{\partial \Phi _{\alpha }}{\partial t}%
\right\rangle $ is the $\alpha $-th band Berry curvature in the momentum and
time spaces,
\begin{equation}
\Phi _{\alpha }=\frac{1}{\sqrt{2\varepsilon _{\alpha }\left( \varepsilon
_{\alpha }-\Delta \right) }}\left(
\begin{array}{c}
-\left( J\cos \frac{qa}{2}+i\delta \sin \frac{qa}{2}\right)  \\
\varepsilon _{\alpha }-\Delta
\end{array}%
\right)   \label{wavefunction}
\end{equation}%
is the eigenwavefunction. The total particle transport along a close loop in
the parameter space can be written as
\begin{equation}
c_{T}=\oint j\left( t\right) dt=\frac{1}{2\pi }\sum_{\alpha =\pm }\oint
dt\int_{-\pi /a}^{\pi /a}f\left( \varepsilon _{\alpha }-\mu \right) \dot{x}%
_{\alpha }\left( q,t\right) dq,  \label{tranatom}
\end{equation}%
where $j\left( t\right) $ is the atom number current, $f\left( \varepsilon
_{\alpha }-\mu \right) $ is the Fermi-Dirac distribution $f\left(
\varepsilon _{\alpha }-\mu \right) =1/\left[ \exp \left( \left( \varepsilon
_{\alpha }-\mu \right) /k_{B}T\right) +1\right] $ for fermionic atoms and
the Bose--Einstein distribution $f\left( \varepsilon _{\alpha }\right) =1/%
\left[ \exp \left( \left( \varepsilon _{\alpha }-\mu \right) /k_{B}T\right)
-1\right] $ for bosonic atoms. For the non-interacting bosonic atoms, all
atoms occupy the lowest energy state and the chemical potential $\mu $ loses
its meaning and total atom transport at the zero temperature becomes
\begin{equation}
c_{T}=\frac{\eta }{a}\oint dt\dot{x}_{-}\left( q_{\min },t\right) ,
\label{tranatom2}
\end{equation}%
where only the lowest energy band has a contribution, $q_{\min }$ is the
quasimomentum at the energy minimum, $\eta $ is the number of atoms in each
double well. The energy minimum of the lowest band $\varepsilon _{-}$
locates at $q_{\min }=0$ except for the parameter $\delta =\pm 1$, where the
band is flat and does not depend on $q$. Note that $\delta =\pm 1$
corresponds to isolated double wells in the lattice (i.e., no inter-well
tunneling), therefore the initial state of the system can be chosen with $q=0
$.

In the case of non-interacting bosonic atoms with an initial $q=0$, the
first term in $\dot{x}_{-}$, $\frac{\partial \varepsilon _{-}}{\partial q}%
|_{q=0}=0$, and the second term $\Omega _{0t}^{-}$%
\begin{equation}
\Omega _{0t}^{-}=\frac{a\delta J}{4}\frac{\partial }{\partial t}\frac{1}{%
\left( J^{2}+\Delta ^{2}-\Delta \sqrt{J^{2}+\Delta ^{2}}\right) }\text{. }
\label{BC}
\end{equation}%
Therefore the total atom transport along the square loop in Fig. 1c is
\begin{equation}
c_{T}=-\frac{\eta }{a}\oint dt\Omega _{0t}^{-}=\eta   \label{tranatom3}
\end{equation}%
when $\Delta \gg J$. When there are $\eta $ atoms in each double well, there
are $\eta $ atoms pumped out the system, agreeing with the numerical results
presented in Figs. 6 and 7.

In the case of strongly interacting bosons, which is equivalent to
non-interacting fermions, the atoms gradually occupy different $q$ states as
the number of filling increases. In the case of the half filling, the lowest
energy band $\varepsilon _{-}$ is fully occupied and the chemical potential $%
\mu $ lies at the gap between two bands. The total transported atoms can be
shown to be $c_{T}=1$ using Eq. (\ref{tranatom}) \cite{Xiao}, which agrees
with the numerical results obtained in Fig 7. In the case of the integer
filling (two atoms per double well), both bands are fully occupied and $%
c_{T}=0$ because the Berry curvatures $\Omega _{qt}^{+}=-\Omega _{qt}^{-}$
for two bands and their contributions to $c_{T}$ cancel with each other.
This result agrees with the numerical results presented in Fig. 6.

Finally, we comment on the validity of using the periodic boundary condition
in the semiclassical theory for the explanation of the numerical results
observed in Sec. III. Eqs. (\ref{Ham4}, \ref{energy}, \ref{tranatom2}) are
introduced by assuming a periodic boundary condition for a single atom and
the fact that the atom occupies the Bloch eigenstate. In general, the
periodic system is not equivalent to our system where the atoms are
initially localized. However, because the initial state we choose is the
product of the state in each isolated double well (no inter-well tunneling),
and the atoms at different double wells do not affect each other (no
interaction), the periodic system can be taken as a multiply copies of our
local system. Therefore the effect of the quantum transport and the
essential physics should be the same for the periodic system and our local
system, which justify the agreement between the semiclassical theory and the
numerical results. The difference between these two systems is that, in a
periodic system, the change of the local atom density after a period cannot
be observed because it is periodic. Instead, what can be observed is the
current, whose integration in one period gives the number of transported
atoms. While in our local system, the transport of atoms is directly
reflected on the density variation because the neighboring lattice sites are
not occupied initially.

\section{Conclusion}

In conclusion, in this paper we study the quantum transport of ultra-cold
bosonic atoms in double well optical lattices where the lattice parameters
can be periodically modulated. The transport of atoms depends strongly on
the atom filling and the interaction. In the strongly interacting region,
the bosonic atoms behave similarly as fermions with quantized transport at
the half filling and no transport at the integer filling. In the weak
interacting region, the quantized transport are robust and the number of the
pumped atoms is proportional to the atom filling per double well. Signature
of the quantum transport of atoms in the momentum distribution is obtained.
In addition to the numerical simulation of the transport dynamics of atoms
in the double well optical lattice using the TEBD algorithm, we develop a
semiclassical model to explain the numerical results in the non-interacting
and strongly interacting limits. Our scheme for the quantized atom transport
does not involve accurate control of the lattice parameters and their time
dependence, thus may provide a robust way for distributing atoms in optical
lattices, which are critically important for quantum computation in optical
lattices as well as the atomtronics applications.

\emph{Acknowledgement} This work is supported by ARO (W911NF-09-1-0248) and
DARPA-YFA (N66001-10-1-4025).

\end{document}